\documentclass[aps,pra,twocolumn,letterpaper,groupedaddress,amsmath,amssymb,longbibliography,10pt]{revtex4-2}
\usepackage{amssymb,amsthm,amsmath,amsfonts}
\usepackage{graphicx,enumerate,bbm,bm}
\usepackage[pdftex,dvipsnames,usenames]{xcolor}
\usepackage[colorlinks=true,linkcolor=blue,citecolor=blue,urlcolor=blue]{hyperref}
\PassOptionsToPackage{hyphens}{url}
\usepackage{braket}
\usepackage{lipsum}

\newcommand{\LA}[1]{\textcolor{black}{#1}}

\newcommand{\TA}[1]{\textcolor{black}{#1}}

\begin{document}

\title{Quantum speedup from nonclassical polarization}

\author{Tim A\ss{}brock}
    \email{tim.assbrock@uni-paderborn.de}
    \affiliation{Theoretical Quantum Science, Institute for Photonic Quantum Systems (PhoQS), Paderborn University, Warburger Stra\ss{}e 100, 33098 Paderborn, Germany}

\author{Jan Sperling}
    \affiliation{Theoretical Quantum Science, Institute for Photonic Quantum Systems (PhoQS), Paderborn University, Warburger Stra\ss{}e 100, 33098 Paderborn, Germany}  

\author{Laura Ares}
    \affiliation{Theoretical Quantum Science, Institute for Photonic Quantum Systems (PhoQS), Paderborn University, Warburger Stra\ss{}e 100, 33098 Paderborn, Germany} 

\date{\today}

\begin{abstract}
We develop a framework for identifying nonclassical speedups in systems with polarization, likewise spin degrees of freedom. 
By confining the dynamics to the manifold of angular momentum coherent states, which act as the classical reference in this case, we compute the speed limit that bounds the rate of change of the state achievable without generating quantum coherence. 
A comparison with the unrestricted quantum speed limit enables the quantitative identification of speedups arising from polarization nonclassicality. 
We apply this framework to the cross-Kerr interaction, demonstrating a persistent speedup scaling as $\mathcal{O}(\sqrt{N})$ with the photon number $N$, \LA{with a parity effect in favor of even photon numbers.}
The results establish polarization nonclassicality as a genuine dynamical resource, linking quantum coherence to quantum-enhanced evolution speeds in nonlinear photonic systems.
\end{abstract}

\maketitle

\section{Introduction}
\label{sec:Introduction}

    Quantum physics is the framework describing physical phenomena at microscopic scales, where superposition, interference, and entanglement are the governing elements determining the structure and evolution of matter and radiation \cite{Nielsen_Chuang_2010,Fabre_2020, Gross_2012, Celi_2016}. 
    Beyond the fundamental insight into nature, quantum mechanics constitutes the foundation for a wide range of research areas, from quantum optics, over condensed-matter physics, to quantum thermodynamics \cite{obrien2009, Wang2020, Noh_2016, Hasegawa2023}.
    In each of these different fields, controlled access to quantum states and their dynamics has not only enabled precision tests of physical laws but also furthered the development of emerging quantum technologies \cite{Preskill2023, Giovannetti2003, benyoucef2024, Ekert1991}. 
    Within this larger landscape, quantum information science has established itself as a central discipline, systematically exploring how quantum resources enable computation, secure communication, and metrology \cite{Nielsen_Chuang_2010, Preskill2023,Shor_1997, grover1996fastquantummechanicalalgorithm}, and providing an operational framework in which the conversion from one quantum resource to another plays a central role.
    
    A key question across the aforementioned domains concerns the impact of quantum resources on the system's dynamical behavior.
    Quantum speed limits (QSLs) formalize this question by bounding the maximal rate at which a system can change between distinct states for a given generator of the evolution, e.g., the Hamiltonian \cite{mandelstam1945, mandelstamm_tamm_PhysRevLett.65.1697, Deffner2017, Giovannetti2003}. 
    These bounds play a dual role: They cast light on structural dynamical constraints in quantum theory and set performance limits for quantum information processing, including, for example, gate execution times, state-transfer protocols, and metrological sensitivity \cite{Deffner2017, Yasmin_2024, Epstein_2017}. 
    The landscape of QSLs stretches across several major directions. 
    First, open-system and perturbed scenarios use QSLs to quantify deviations from reference trajectories and directly relate to quantum Fisher information and environmental timescales \cite{Pires2024, Yadin2024}. 
    Second, thermodynamic and stochastic formulations now bring QSLs and thermodynamic uncertainty relations in a unifying picture \cite{Hasegawa2023}. 
    Third, bounds for observables constrain the rate of change of expectation values, being important for entanglement generation, modular Hamiltonians, and quantum batteries \cite{Shrimali2024, Shrimali2025}. 
    Non-Hermitian and measurement-induced dynamics introduce gain and loss channels and means for continuous monitoring into the QSLs \cite{Nishiyama2025}, and anomalous transport and memory effects drive fractional-time generalizations for open systems \cite{Wei2023}. 
    At the many-body level, operator-flow and correlation-spreading QSLs complement Lieb-Robinson-type bounds by characterizing limits to scrambling and information propagation \cite{Hinrichs_2024, Carabba2022}. 
    A recurring theme across these developments is that genuinely quantum resources can accelerate thermodynamic processes, control operations, and information flow \cite{Luo2024}.
    
    The promising field of photonic quantum technologies provides a particularly suitable context to investigate the quantum aspect of the dynamics because of low-loss transmission, resilience against decoherence, mature state-preparation techniques, and compatibility with existing optical infrastructure \cite{benyoucef2024, obrien2009, Wang2020}.
    Polarization encoding, in particular, allows high-fidelity manipulation and interferometric stability \cite{Caves1981,LIGO_nphys2083},  naturally interfacing with multiphoton interference and entanglement generation \cite{horodecki2009_RevModPhys.81.865,Ekert1991}.
    Moreover, recent theoretical and experimental work has further established a strict equivalence between polarization nonclassicality and multiphoton entanglement, thereby unifying two key signatures of quantumness within a single operational framework \cite{ares2024photonicentanglementpolarizationnonclassicality}. 
    Thus, polarization photonic systems offer a platform where dynamical advantages induced by nonclassical resources can be quantitatively assessed via experimentally accessible observables. Yet, a theoretical analysis and methodology applicable to experiments in nonlinear polarization optics is missing to date.

    In this work, we investigate how quantum resources manifest as dynamical advantages by comparing the temporal behavior of polarization (likewise, spin) systems to the classical scenario in which nonclassical resources are not accessible.
    To this end, we derive equations of motion that confine the dynamics to the manifold of classical polarization states, the so-called angular momentum coherent states (AMCSs), preventing the buildup of quantum coherence at all times \cite{Sperling_cl_ev_in_qs}. 
    AMCSs are a suitable classical reference since, being $\mathrm{SU}(2)$-coherent states, minimize angular momentum uncertainties and exhibit shot-noise-limited Stokes fluctuations, and directly relate to the coherent states of the harmonic oscillator \cite{amcs_2_luis_PhysRevA.84.042111,amcs_1_atkins_dobson,amcs_9_Perelomov_1977}.
    Furthermore, we compute the speed limits for the restricted dynamics, allowing for a direct comparison with the actual QSLs of the process and thereby assessing the advantages of nonclassical processes.

    As an application, we apply our method to investigate the cross-Kerr effect, which arises from the third-order nonlinear susceptibility of the medium that leads to an intensity-dependent refractive index~\cite{BoydNLO}. 
    Cross-Kerr nonlinearities play a central role in quantum optics and quantum information science \cite{cross_kerr_h_PhysRevA.91.043822,Khorasani_2019,schroedinger_cat_states_He2023}, enabling quantum nondemolition measurements \cite{Fushman_2007}, the realization of photonic controlled-phase gates \cite{cphase_PhysRevA.94.023833,Luo2016}, and facilitating the generation of nonclassical states \cite{schroedinger_cat_states_He2023}. 
    More broadly, cross-Kerr interactions represent a fundamental mechanism for implementing all-optical quantum logic operations \cite{PhysRevLett.93.250502}. 
    See Refs. \cite{PhysRevA.64.023805,RevModPhys.77.633,doi:10.1073/pnas.1524117113,PhysRevLett.111.053601} for experimental applications.
    Our findings unambiguously link the quantum coherence generated by this nonlinear interaction to the speedup of the dynamics.

\section{Preliminaries}
\label{sec:Preliminaries}
    To establish a clear foundation for the forthcoming derivation, we briefly review some required concepts in this section.
    We begin by revisiting the notions of quantum coherence and AMCSs.
    Then, we summarize the procedure that restricts the dynamics to classical states, and specify the definition of the specific QSL for dynamical polarization nonclassicality.

    \subsection{Quantum coherence}
    \label{sec:preliminaries_quant_coh}
        We here adopt the resource-theoretic notion of quantum coherence presented in Ref. \cite{Quant_coherence_Baumgratz_2014}. 
        Throughout, quantum coherence is defined relative to a fixed set of reference states $\ket{\psi(\mathbf{q})}$, acting as the classical reference, where the tuple $\mathbf{q}$ serves as the parametrization for such states \cite{Sperling_cl_ev_in_qs}. 
        A quantum state is then said to exhibit quantum coherence if it cannot be written as a convex mixture of these basis states.
        Within this framework, quantum coherence serves as the marker of nonclassicality, which can, for example, translate into the quasiprobability distribution representing the state showing negative values \cite{Sperling2018, Sperling2020a,Sperling2025,Bohmann2020,Tan2020,Linowski2024}.  
        For instance, this notion of quantum coherence encompasses quantum effects like entanglement by choosing product states as the basis states.

    \subsection{Angular momentum coherent states}
        Now, we elaborate on the set of classical reference states for polarization.
        AMCSs are built on the Schwinger representation \cite{amcs_4_schwinger_osti_4389568}, introducing two independent bosonic modes with the creation operators $\hat{a}_\pm^\dagger$ representing two orthogonal polarizations.
        These operators satisfy the canonical commutation relations and yield the angular momentum operators
        \begin{equation}\label{ch3.1_ang_mom_operators}
            \hat{J}_\pm=\hat{a}_\pm^\dagger\hat{a}_\mp=\hat{J}_{x}\pm i\hat{J}_{y}\quad\text{and}\quad \hat{J}_z=\frac{1}{2}\left(\hat{n}_+-\hat{n}_-\right),
        \end{equation}
        with $\hat{n}_\pm=\hat{a}^\dagger_\pm\hat{a}_\pm$ \cite{amcs_4_schwinger_osti_4389568, amcs_2_luis_PhysRevA.84.042111}. The above operators form the ladder operators and spin-$z$ operator for an angular momentum algebra, represented by the corresponding Stokes operators in quantum optics.
        In two-mode Fock basis $\{\ket{n_+,n_-}: n_\pm\in\mathbb N\}$, the AMCS $\ket{s_N}$ for a $N$-photon state with $n_+$ photons in the ``$+$" mode and $n_-=N-n_+$ photons in the ``$-$" mode is given by \cite{ares2024photonicentanglementpolarizationnonclassicality}
        \begin{equation}
        \label{preliminaries:AMCS}
            \ket{s_N}=\sum_{n_+=0}^N{N\choose n_+}^\frac{1}{2}\alpha_+^{n_+}\alpha_-^{n_-}\ket{n_+,n_-},
        \end{equation}
        with $|\alpha_+|^2+|\alpha_-|^2=1$. 
        In contrast to Glauber coherent states, AMCSs for $N$ photons are not eigenstates of the annihilation operators $\hat{a}_\pm$, but one can show that said operators act as
        \begin{equation}\label{eq:AnnihilationIdentity}
            \hat{a}_\pm\ket{s_N}=\sqrt{N}\alpha_\pm\ket{s_{N-1}}\text{,}
        \end{equation}
        effectively mapping the $N$-photon AMCS onto the corresponding $(N-1)$-photon state. 
    
    \subsection{Classical evolution}
    \label{sec:preliminaries:cl_ev}
        The definition of classical evolution considered here forbids the presence of quantum coherence for all times $t$ \cite{Sperling_cl_ev_in_qs}, in contrast with the common interpretation encompassing processes that map classical states into classical states via a Kraus operator after a certain time $T$.
        Based on the notion for the continuum of times, one can derive restricted equations of motion for a general process by imposing on the evolution the constraint to remain classical, which yields
        
        \begin{equation}
        \label{preliminaries:cl_eqn_of_motion}
            \bra{ \nabla_{\boldsymbol{q}}\psi(\boldsymbol{q}) }\left(i\hbar\partial_t\ket{\psi(\boldsymbol{q})}-\hat{H}\ket{\psi(\boldsymbol{q})}\right)
            =\boldsymbol{0}.
        \end{equation}
        Note that the vector $\ket{\nabla_\mathbf{q}\psi(\mathbf{q})}=\nabla_\mathbf{q}\ket{\psi(\mathbf{q})}$ is the derivative of the classical reference state with respect to its parametrization $\mathbf{q}$, i.e., tangent vectors to the manifold of classical states $|\psi(q)\rangle$ at point $\boldsymbol{q}$.
        Thus, Eq. \eqref{preliminaries:cl_eqn_of_motion} describes the common Sch\"odinger evolution projected onto the classical states.
        A comparison with the dynamics described by the actual -- i.e., nonprojected -- Schr\"odinger equation allows us to isolate the temporal effect of nonclassicality in the evolution of states.

    \subsection{Quantum speed limit}
        Finally, we recall the concept of QSL, which characterizes the maximal rate of change of a quantum system evolving between two distinct states. 
        Note that various formulations exist, e.g., the early Mandelstamm-Tamm bound and the Margolos-Levitin bound \cite{mandelstam1945, mandelstamm_tamm_PhysRevLett.65.1697, Deffner2017, Yasmin_2024}. 
        In this work, we utilize the geometric approach to QSLs, where the instantaneous speed of evolution is identified with the trace norm of the rate of change of the density operator, which quantifies the infinitesimal change in state distinguishability \cite{Deffner2013,Deffner2013a,Deffner2017}. 
        
        Explicitly, for a quantum state $\hat{\rho}=\ket{\Psi}\bra{\Psi}$, the rate of change is given by $\left\lVert\partial_t\ket{\Psi}\bra{\Psi}\right\lVert_1$, where $\left\lVert\cdot\right\lVert_1$ is the trace norm, equivalently defined as the sum of the modulus of the eigenvalues of a self-adjoint operator. 
        The QSL is then the maximal rate of change of the system \cite{Deffner2013,Deffner2017},
        \begin{equation}\label{preliminaries:QSL}
            \text{QSL}=\sup\limits_{\ket{\Psi}}\left\lVert\partial_t\ket{\Psi}\bra{\Psi}\right\lVert_1=\sup\limits_{\ket{\Psi}}\left\lVert\frac{1}{i\hbar}\left[\hat{H},\ket{\Psi}\bra{\Psi}\right]\right\lVert_1.
        \end{equation}
        Note that, in closed systems, we can limit the study to pure states because of convexity.
        The rate of change can be directly linked to the dynamics' generator, the Hamiltonian $\hat{H}$, and thus to the energy of the system, through the von Neumann equation \cite{Yasmin_2024}. 
        The QSL for the quantum dynamics without any constraint on the evolution results in
        \begin{equation}\label{preliminaries:QSL_Energy}
            \text{QSL}=\frac{E_{\text{max}}-E_{\text{min}}}{\hbar},
        \end{equation}
        where $E_{\text{max}}$  and $E_{\text{min}}$ denote the supremum and infimum over $\ket{\Psi}$ of the energy values $E=\bra{\Psi}\hat{H}\ket{\Psi}$, respectively;
        see, for example, Ref. \cite{Yasmin_2024} and references therein for detailed derivations.
        
        \LA{Importantly, }\TA{for the pure-state unitary dynamics considered here, the trace norm rate of change is inspired by the standard Mandelstam-Tamm geometric speed, as well as the speed associated with the quantum Fisher information \cite{Taddei_2013,Pires_2016,Deffner2017}. 
        }\LA{Specifically, } \TA{the quantum Fisher information associated with estimating the time $t$ is \cite{Braunstein_1994,paris2009quantumestimationquantumtechnology}}
        \begin{equation}
        \LA{F_Q(t)=4\left(\langle\dot\psi|\dot\psi\rangle-|\langle\psi|\dot\psi\rangle|^2\right) =\|\dot\rho\|_1^2},
        \end{equation}
        \TA{where the dot notation, $|\dot\psi\rangle=\partial_t|\psi\rangle$, was used. Equivalently, the Fubini-Study velocity is $v_{\rm FS}=\sqrt{F_Q(t)}/2=\|\dot\rho\|_1/2$.
        Under unitary evolution, this becomes $\|\dot\rho\|_1=2\Delta_\psi \hat H/\hbar$, where $\Delta_\psi \hat H$ is the energy uncertainty of $\ket{\psi}$. This corresponds to the Mandelstam-Tamm speed up to the conventional factor of $2$ between trace distance and Fubini-Study angle velocities. 
        Maximization over pure states gives $(E_{\max}-E_{\min})/\hbar$, which is the expression used in Eq. \eqref{preliminaries:QSL_Energy}. 
        The extremal state realizing this maximum is the equal superposition of the lowest and highest energy eigenstates and, for $E_{\min}=0$, also saturates the Margolus-Levitin orthogonalization bound \cite{Margolus_1998}. 
        Thus, for the fixed-$N$ closed-system dynamics, the trace norm, quantum Fisher information, Mandelstam-Tamm, and optimized Margolus-Levitin perspectives yield the same spectral-range scaling.}
    
\section{Restricted framework}
\label{sec:derivation_polarization}
    In this section, we restrict the dynamics of a general Hamiltonian to classical states for polarization systems and compute the thereby restricted speed limit. 
    This constitutes a general framework, allowing for the certification of polarization-nonclassical processes, which then can be applied to concrete processes (see the next section).

    \subsection{Equations of motion}
        We derive the equations of motion constrained to AMCSs by introducing $\ket{s_N}$ in the Schr\"odinger-type equation of motion [Eq. (\ref{preliminaries:cl_eqn_of_motion})].
        According to Eq. \eqref{preliminaries:AMCS}, we identify the classical parametrization as $\boldsymbol{q}=\boldsymbol{\alpha}=(\alpha_+,\alpha_-)\in\mathbb C^2$, being the only varying parameters in $\ket{s_N}$.
        Thus, the time derivative of the AMCS takes the form
        \begin{align}\label{preliminaries_time_deriv}
            \partial_t{\ket{s_N}}=\sqrt{N}\left(\dot{\alpha}_+\hat{a}_+^\dagger+\dot{\alpha}_-\hat{a}_-^\dagger\right)\ket{s_{N-1}},
        \end{align}
        where the identity in Eq. \eqref{eq:AnnihilationIdentity} was used.
        Further, we can rewrite $\bra{\nabla_\mathbf{\alpha}s_N}\hat{H}\ket{s_N}=\nabla_\mathbf{\alpha^*}\bra{s_N}\hat{H}\ket{s_N}$ since neither $\hat{H}$ nor $\ket{s_N}$ depends on $\alpha_\pm^*$. 
        Substituting the above expressions into Eq. \eqref{preliminaries:cl_eqn_of_motion}, we obtain the following coupled equations of motion:
        \begin{equation}\label{derivation:class_eqn}
            \begin{pmatrix}\dot{\alpha}_+\\ \dot{\alpha}_-\end{pmatrix}=-\frac{i}{\hbar N^2}\mathbf{\Lambda}\begin{pmatrix} \partial_{\alpha_+^*}\langle\hat{H}\rangle \\ \partial_{\alpha_-^*}\langle\hat{H}\rangle \end{pmatrix},
        \end{equation}
        for each component of $\boldsymbol\alpha=(\alpha_+,\alpha_-)$ and with $\langle\hat{H}\rangle=\bra{s_N}\hat{H}\ket{s_N}$ and the $\boldsymbol\alpha$-dependent matrix 
        \begin{equation}
        \label{eq:lambda}
            \mathbf{\Lambda}=\begin{pmatrix}1+(N-1)|\alpha_-|^2 & -(N-1)\alpha_-^*\alpha_+\\-(N-1)\alpha_+^*\alpha_- & 1+(N-1)|\alpha_+|^2\end{pmatrix}.
        \end{equation}
        These equations of motion formally resemble Hamilton's equations in classical mechanics, but with respect to a quantum-state-dependent symplectic structure that arises from the restriction to the manifold of AMCSs \cite{CannasDaSilva2001}.
        Moreover, Eqs. (\ref{derivation:class_eqn}) and (\ref{eq:lambda}) can be equivalently reformulated in terms of the normalized, classical Stokes vector 
        \begin{equation}\label{norm_Bloch_vec}
            \vec{r}=\frac{2}{N}\braket{\vec{J}}=\frac{2}{N}\begin{pmatrix}
                \text{Re}\braket{\hat{J}_+}\\\text{Im}\braket{\hat{J}_+}\\\braket{\hat{J}_z}
            \end{pmatrix}=\begin{pmatrix}r_x\\r_y\\r_z
            \end{pmatrix},
        \end{equation}
        where the expectation values are taken with respect to $\ket{s_N}$.
        Then, Eq. \eqref{derivation:class_eqn} takes the form of a closed dynamical equation of Lie-Poisson type \cite{Marsden1999,Marsden1999a,Holm1999}; that is, by defining the vector
        \begin{equation}
            \vec{h}=\begin{pmatrix}
                \partial_{r_x} \langle\hat{H}\rangle\\
                \partial_{r_y} \langle\hat{H}\rangle\\
                \partial_{r_z} \langle\hat{H}\rangle
            \end{pmatrix}=\nabla_{\vec{r}}\langle\hat{H}\rangle,
        \end{equation}
        we can rewrite Eq. (\ref{derivation:class_eqn}) as
        \begin{equation}
            \dot{\vec{r}}=-\frac{2}{\hbar N}\vec{r}\times\vec{h}.
        \end{equation}
        In this representation, the restricted evolution on the AMCS manifold interestingly takes the form of a classical rigid-body equation, also known as Euler's rotation equations.
        This formulation makes the geometric structure of the classical evolution explicit and allows for a direct comparison with classical angular momentum and polarization dynamics.

        \TA{
        The restricted evolution defined here 
        acts as a resource-theoretic benchmark, rather than as a phenomenological model of classical 
        experiments. 
        The purpose of the AMCS projection is therefore to 
        }
        \LA{ identify the closest evolution within the classical states, isolating and gauging the impact that nonclassical polarization has on the actual evolution, and not necessarily to capture any particular classical evolution in experiment.}
        \TA{In realistic settings, technical noise, loss,
        and statistical fluctuations may lead to open-system dynamics.}
        \LA{Within our framework, this scenario can be addressed by constraining the corresponding Lindblad equation instead of the Schrödinger equation in Eq. (\ref{preliminaries:cl_eqn_of_motion}).
        A similar approach has already been utilized to analyze the impact of entanglement in open systems in Refs. \cite{pinske2024separabilitylindbladequationdynamical,ares2024restrictedMonteCarlowave-functionmethod}}.
    
    \subsection{Classical quantum speed limit}
        Now, we introduce the speed limit for the evolution governed by the restricted equations of motion in Eq. (\ref{derivation:class_eqn}).
        We refer to the resulting quantity as the classically restricted quantum speed limit ($\text{QSL}_{\text{cl}}$). 
    
        In order to evaluate the trace norm in Eq. (\ref{preliminaries:QSL}), we express $\partial_t\ket{s_N}=c_\parallel\ket{s_N}+c_\perp\ket{s_N^\perp}$ in the orthogonal basis $\{\ket{s_N}, \ket{s_N^{\perp}}\}$, accounting for the parallel and orthogonal components.
        This decomposition, together with the time derivative of $|s_N\rangle$ in Eq. (\ref{preliminaries_time_deriv}), allows us to write $\partial_t(\ket{s_N}\bra{s_N})$ as a matrix with eigenvalues $\pm|\lambda|=\pm\sqrt{N}|\dot{\alpha}_+\alpha_--\dot{\alpha}_-\alpha_+|$. 
        Thus, the trace norm that defines $\text{QSL}_{\text{cl}}$ reads as 
        \begin{equation}\label{deriv:qsl_cl_c_perp}
            \text{QSL}_{\text{cl}}=\sup\limits_{\ket{s_N}}2|\lambda|=2\sqrt{N}\sup\limits_{\ket{s_N}}|\dot{\alpha}_+\alpha_--\dot{\alpha}_-\alpha_+|.
        \end{equation}
        Substituting the classical equations of motion from Eq. (\ref{derivation:class_eqn})
        into this expression ultimately yields the maximal rate of change of a polarization system when no quantum coherence is allowed to build up during the evolution,
        \begin{equation}\label{derivation:qsl_cl_final}
            \text{QSL}_{\text{cl}}=\frac{2}{\hbar\sqrt{N}}\sup\limits_{\ket{s_N}}\left\vert\alpha_-\frac{\partial\langle\hat{H}\rangle}{\partial{\alpha_+^*}}-\alpha_+\frac{\partial\langle\hat{H}\rangle}{\partial{\alpha_-^*}}\right\vert.
        \end{equation}
        In the context of a classical, analytical mechanics, note that the quantity over which the supremum is taken plays the role akin to an angular momentum.
        Furthermore, the result enables us to compare the unrestricted QSL [Eq. (\ref{preliminaries:QSL_Energy})], with its restricted counterpart $\text{QSL}_{\text{cl}}$ from Eq. (\ref{derivation:qsl_cl_final}). 
        In addition, we certify a nonclassicality-enabled speedup when $\text{QSL}>\text{QSL}_{\text{cl}}$ holds true, and the difference of both quantities quantifies the amount of quantum speedup.

    \color{black}
    \subsection{Certification criteria}
    To construct measurable criteria to certify the presence of nonclassicality-assisted speedup, we consider the expected value of an observable $\hat{L}$ at two different points in time,
    \begin{equation}
    \label{eq:diference_espected_values}
    \begin{split}
        |\langle\hat L\rangle(t_2)-\langle\hat L\rangle(t_1)|=&|\mathrm{tr}\{[\hat\rho(t_2)-\hat\rho(t_1)]\hat{L}\}|\\
        \leq &\|\hat\rho(t_2)-\hat\rho(t_1)\|_1\cdot\|\hat L\|_\infty,
    \end{split}
    \end{equation}
    using the H\"older inequality in the second row, where $\|\cdot\|_\infty$ is the spectral norm.
    We can now translate our limit for the instantaneous velocity $\|\partial_t(\ket{s_N}\bra{s_N})\|_1=\|\partial_t\hat\rho_\mathrm{cl}(t)\|_1\leq \mathrm{QSL}_\mathrm{cl}$ to bound the change rate on a finite time interval \cite{Yasmin_2024}, $\|\hat\rho_\mathrm{cl}(t_2)-\hat\rho_\mathrm{cl}(t_1)\|_1\leq (t_2-t_1) \mathrm{QSL}_\mathrm{cl}$.
    Substituting in Eq. (\ref{eq:diference_espected_values}), the difference in the expected value of an observable when no coherence is allowed becomes limited,
    \begin{equation}
        |\langle\hat L\rangle_\mathrm{cl}(t_1)-\langle\hat L\rangle_\mathrm{cl}(t_2)|\leq (t_2-t_1) \mathrm{QSL}_\mathrm{cl}\|\hat L\|_\infty.
    \end{equation}
    As a consequence, any measurement beyond this limit certifies the presence of nonclassicality imprinted on a speedup of the process. 
    The observable $\hat L$ can be chosen according to the process under consideration and the experimentally accessible measurement devices.
    \color{black}
\section{Application: Cross-Kerr effect}
\label{sec:application}
    
    In this section, we apply the methods developed above to the cross-Kerr effect, which, for two quantized modes, is described by the Hamiltonian
    \begin{equation}\label{ck:hamiltonian}
        \hat{H}_\text{int} = \varepsilon\, \hat{n}_+ \hat{n}_-,
    \end{equation}
    where $\hat{n}_\pm = \hat{a}_\pm^\dagger \hat{a}_\pm$ are the photon-number operators and $\varepsilon$ denotes the nonlinear coupling strength \cite{cross_kerr_h_PhysRevA.91.043822,Khorasani_2019}. 
    This interaction gives rise to nontrivial phase dynamics due to the coupling between the modes. 
    We adopt natural units ($\hbar = 1$) and omit the free evolution of the modes (i.e., $\hat{H}=\hat{H}_\text{int}$) because a free evolution $\propto \hat n_++\hat n_-$ always trivially maps AMCSs to AMCSs\LA{, creating no coherence and having no impact on the difference between classical and quantum dynamics.}
    \LA{In addition, note that an ideal lossless scenario is considered. }
    \TA{Therefore, the following analysis} \LA{represents a} \TA{ closed-system benchmark. 
    In realistic media, effects like absorption, scattering, dephasing, mode mismatch, and photon loss can reduce the }\LA{ coherence in the system and the associated speedup}\TA{, especially for large $N$.}
    \LA{To address the lossy scenario within our framework, an open-system approach can be introduced, analogously to Refs. \cite{pinske2024separabilitylindbladequationdynamical,ares2024restrictedMonteCarlowave-functionmethod}. 
    This is based on the same idea of classically restricted dynamics as introduced here.}
    
    \subsection{Dynamics}
        The full quantum evolution can be computed via the Heisenberg equation for the field operators, $\partial_t\hat{a}_\pm=-i(1+\varepsilon\hat{n}_\mp)\hat{a}_\pm$.
        Therefore, up to global phases, the unrestricted evolution of an initial AMCS is given by
        \begin{equation}
            \ket{\sigma_N(t)}\!=\!\sum_{n_+=0}^N{\!N\!\choose \!n_+\!}^\frac{1}{2}\alpha_{+}^{n_+}\alpha_{-}^{n_-}e^{-i\varepsilon n_+ n_-t}\ket{n_+,n_-},
        \end{equation}
        with $\alpha_\pm=\alpha_{\pm}(0)$ defining the initial parameters of the AMCS. Importantly, we note that the nonlinear phase of each component $e^{-i\varepsilon n_+ n_-t}=e^{-i\varepsilon (N n_+-n_+^2)t}$ drives the state out of the manifold of AMCSs.

        To derive the restricted dynamics, we compute and insert $\braket{\hat{H}}=\varepsilon N(N\!-\!1)|\alpha_+(t)|^2|\alpha_-(t)|^2$ into Eq. (\ref{derivation:class_eqn}), leading to the differential equations
        \begin{equation}
        \begin{split}
            \dot{\alpha}_\pm(t) =& -iA
            |\alpha_\mp(t)|^2\alpha_\pm(t)\\
            &\times\left[1+(N\!-\!1)(|\alpha_\mp(t)|^2-|\alpha_\pm(t)|^2)\right],
        \end{split}
        \end{equation}
        with $A=\varepsilon (N\!-\!1)/N$. 
        Solving these equations reveals the constants of motion $\lvert\alpha_\pm(t)\rvert=\lvert\alpha_\pm(0)\rvert=\lvert\alpha_\pm\rvert$ and thus allows us to write the solution as
        \begin{equation}
            \alpha_\pm(t)=\alpha_\pm (0) e^{-i\omega_\pm t},
        \end{equation}
        with $\omega_\pm=A|\alpha_\mp|^2\left[1\!+\!(N\!-\!1)(|\alpha_\mp|^2\!-\!|\alpha_\pm|^2)\right]$.
        Thus, the classical evolution of the AMCSs consists of a rotation in the complex plane with a constant radius,  
        \begin{equation}
            \ket{s_N(t)}\!=\!\sum_{n_+=0}^N{\!N\!\choose \!n_{+}\!}^\frac{1}{2}\alpha_{+}(0)^{n_+}\alpha_{-}(0)^{n_-}e^{-in_+\tilde\omega t}\ket{n_+,n_-},
        \end{equation}
        where $\tilde\omega=\omega_+-\omega_-=\varepsilon(N-1)(|\alpha_-|^2-|\alpha_+|^2)$. 
        We now analyze the fundamental difference between the fully quantum and the AMCS-restricted trajectories by visualization on the Poincar\'e sphere, describing the normalized Stokes vector $\vec r$ in Eq. \eqref{norm_Bloch_vec}.
        Note that this corresponds to the Bloch sphere for spin-$1/2$ systems.
\begin{figure}[!t]
    \centering
    \includegraphics[width=1\linewidth]{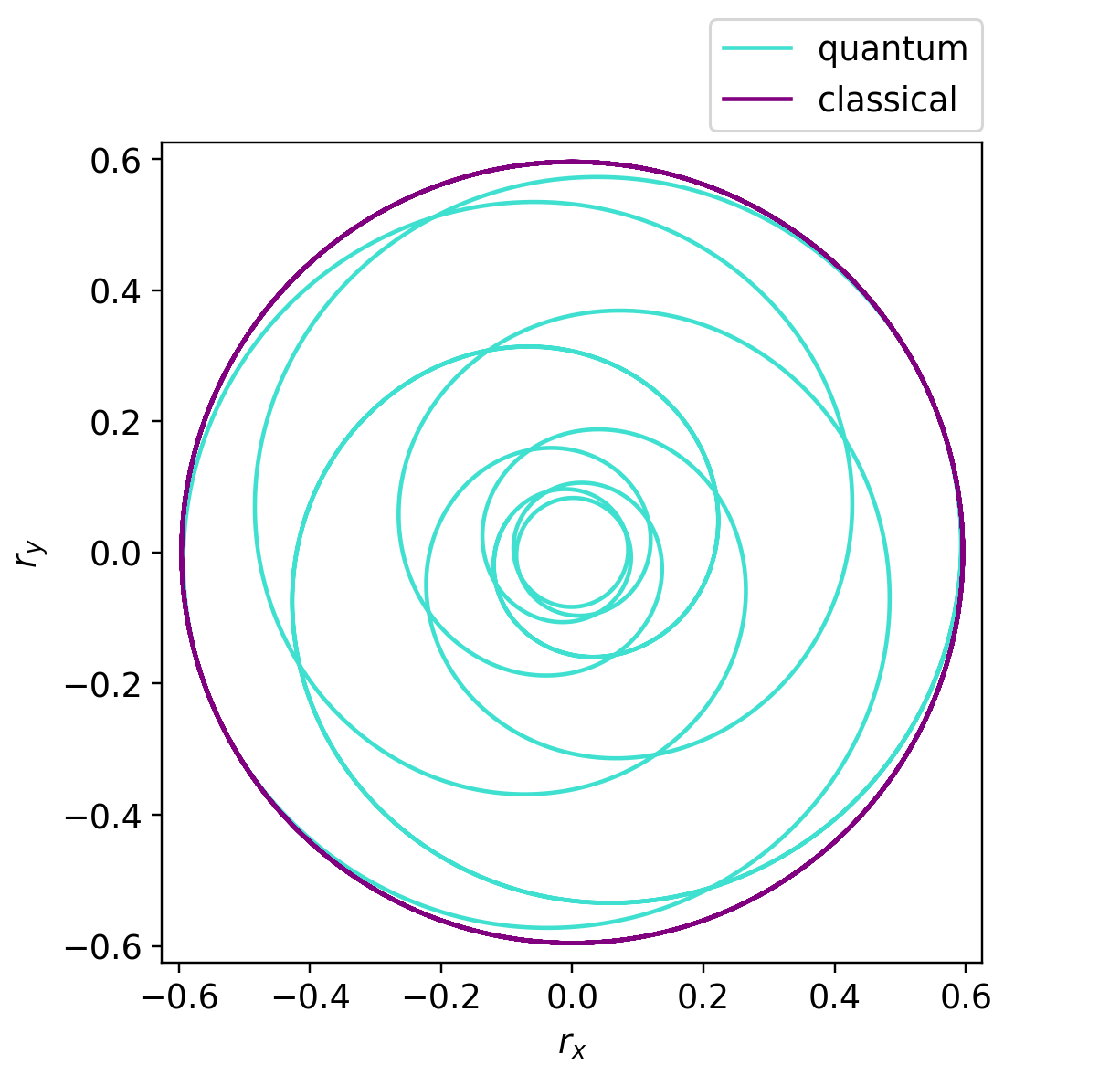}
    \caption{Time evolution of the normalized Stokes vectors for the unrestricted (turquoise) and classical (purple) case for $\varepsilon=0.05$ and the initial conditions $\alpha_{+}=0.95$ and $ \alpha_{-}=0.29+0.12i$.
    Note that the polar angle of the evolution on the Poincar\'e sphere is constant, and we thus only show the $r_x-r_y$ plane.}
    \label{fig:application:bloch_disks}
\end{figure}
        
        In Fig. \ref{fig:application:bloch_disks}, the classically evolved state remains on the boundary of the disk that intersects the Poincar\'e sphere with the $r_x-r_y$ plane, proving that it stays classical for all times $t$. In contrast, the unrestricted state erratically oscillates within the intersecting disk during the evolution, hinting at the presence of quantum coherence, which is caused by the aforementioned, highly nonlinear phases.
\begin{figure}[b]
    \centering
    \includegraphics[width=1\linewidth]{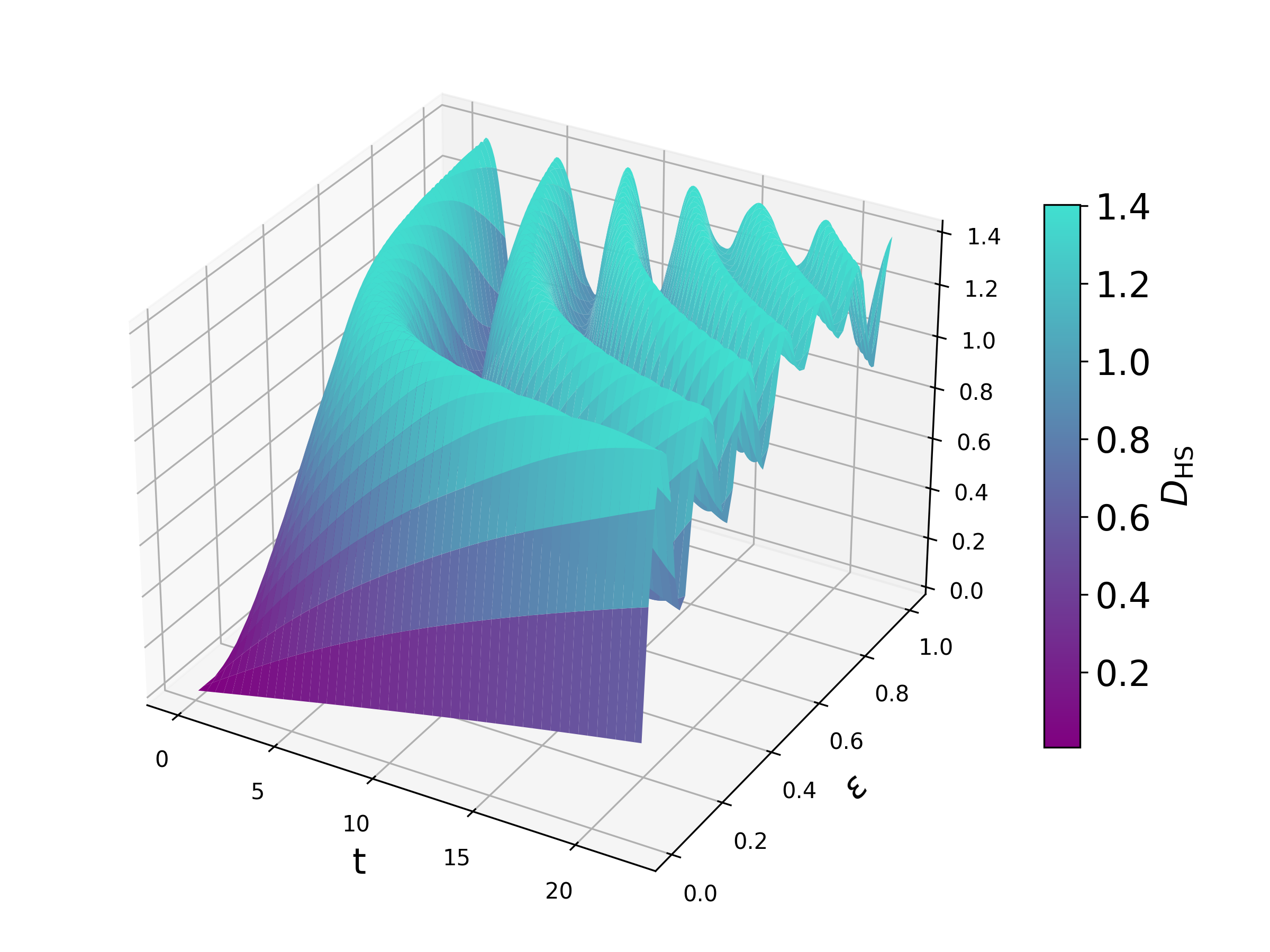}
    \caption{Hilbert-Schmidt distance $D_{\text{HS}}$ between classical and quantum evolutions as a function of  time $t$ and coupling strength $\varepsilon$ for $|\alpha_+|^2=0.9$ and $N=10$.}
    \label{fig:application_dhs_1}
\end{figure}        
        To analyze the impact of the coupling strength $\varepsilon$ on the difference of the classical and quantum dynamics of the states, we compute the Hilbert-Schmidt distance \cite{d_hs_PhysRevA.84.032120} between both evolved density operators, which reads as
        \begin{equation}
            D_{\text{HS}}(t)
            =
            \sqrt{2}\sqrt{1-\lvert\braket{s_N(t)\vert \sigma_N(t)}\rvert^2}
        \end{equation}
        for pure states. Figure \ref{fig:application_dhs_1} shows $D_{\text{HS}}$ as a function of $\varepsilon$ and $t$, revealing an oscillating structure that arises from the periodicity of both evolutions. 
        The stronger the coupling, the closer $ D_{\text{HS}}(t)$ oscillates near the upper limit of $\sqrt{2}$, which corresponds to perfect orthogonality between the classically evolved and quantum evolved states. 
        For weak coupling, however, the states remain similar [$D_{\text{HS}}(t)\approx 0$] for longer times, as one would expect.
        Thus, we conclude that the cross-Kerr interaction introduces a nonclassical polarization evolution to the system.

    \subsection{Speed limits}
        To obtain the classical speed limit for a cross-Kerr interaction, we can directly substitute $
        \bra{s_N}\hat{H}\ket{s_N}
        $ into Eq. (\ref{derivation:qsl_cl_final}), which gives us the following expression:
        \begin{equation}\label{application:qsl_cl_calculation}
            \text{QSL}_{\text{cl}}=2\varepsilon\sqrt{N}(N\!-\!1)\sup\limits_{\alpha_\pm}\left\vert\alpha_+\alpha_-\left(\lvert \alpha_-\rvert^2\!-\!\lvert\alpha_+\rvert^2\right)\right\vert.
        \end{equation}
        We can straightforwardly solve this optimization problem by setting $|\alpha_+|^2=p$ and $|\alpha_-|^2=1-p$ and searching for the maximum for $p\in[0,1]$, which is $1/4$. 
        Thus, the maximum rate of change of a state restricted to the manifold of AMCSs in a cross-Kerr medium is explicitly given by
        \begin{equation}
         \label{eq:QSL_cl}   \text{QSL}_{\text{cl}}=\frac{\varepsilon\sqrt{N}(N-1)}{2}.
        \end{equation}
        Importantly, this limit scales linearly with $\varepsilon$ and as $\mathcal{O}(N^{3/2})$ with the photon number $N$.

        In contrast, the unrestricted QSL in Eq. (\ref{preliminaries:QSL_Energy}) is evaluated through the eigenvalues of the Hamiltonian for a fixed $N$, $E(n_+)=\varepsilon n_+(N-n_+)$, with $ n_+\in[0,N]$.
        Thus, we find the ultimate speed limit without restrictions as
        \begin{equation}
        \label{application:Q_N}
            \text{QSL}=\varepsilon\frac{N^2-(N~\text{mod}~2)}{4}.
        \end{equation}
        
\begin{figure}[t]
    \centering
    \includegraphics[width=1\linewidth]{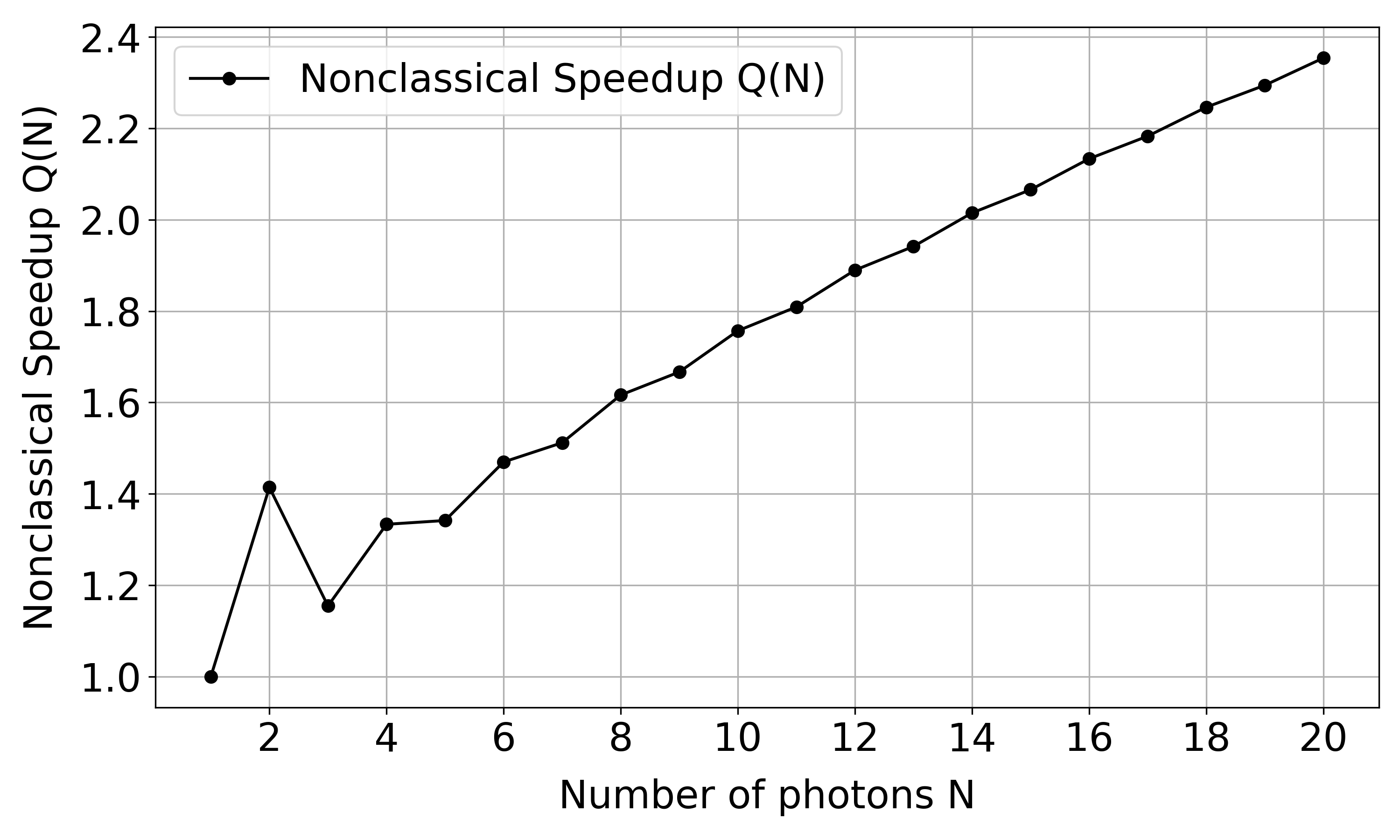}
    \caption{Plot of the quantum speedup via the ratio $Q(N)$ of the unrestricted and the classical QSLs. Since $Q(N)>1$, the nonclassical speedup of cross-Kerr processes is certified. The plot exhibits an advantage of even photon numbers for small photon numbers $N$ and scales as $\mathcal{O}(N^{1/2})$ for larger $N$.}
    \label{fig:application:Q_N}
\end{figure}

        To quantitatively compare Eqs. (\ref{eq:QSL_cl}) and (\ref{application:Q_N}), it is convenient to define the ratio
        \begin{equation}\label{eq:KerrGrowth}
            Q(N)=\frac{\text{QSL}}{\text{QSL}_{\text{cl}}}=\frac{N^2-(N~\text{mod}~2)}{2\sqrt{N}(N-1)}.
        \end{equation}
        The amount to which $Q(N)>0$ holds true certifies the speed gained through the quantum part of the evolution.
        Here, $Q(N)$ is strictly bigger than 1 ($\forall N>1$), verifying a nonclassical speedup of optical cross-Kerr processes because of quantum coherence effects. 
        Figure \ref{fig:application:Q_N} shows the nonclassical speedup $Q(N)$, in which we can easily see the growth of $Q(N)$ as $\mathcal{O}(N^\frac{1}{2})$ [cf. Eq. \eqref{eq:KerrGrowth}].
        Compared to the previous analysis for specific initial states, the QSLs used here are independent of this initial choice.
        Further, the ratio $Q(N)$ is even independent of the coupling strength (except for $\varepsilon=0$, which would cause a division by zero) and thus represents a universal speedup of a cross-Kerr nonlinearity.
        Interestingly, we also observe the difference for odd and even $N$, which is more pronounced for small $N$. 
        \TA{Even $N$ allow the cross-Kerr spectrum to attain its maximal value exactly at $n_+=N/2$, whereas for odd $N$, the maximum is shifted to the two neighboring photon-number sectors. 
        Consequently, even photon numbers provide a slightly larger speedup ratio. 
        This observation may be relevant for few-photon quantum-information protocols, where the choice of photon number can affect the achievable gate speed in nonlinear photonic implementations.}

\section{Conclusion}
\label{sec:Conclusion}

    In this work, we developed a general framework to identify nonclassical speedups in polarization (likewise, angular momentum) systems.
    First, we derived restricted equations of motion, which render it possible to describe the dynamics of the polarization states confined to the manifold of classical AMCSs, thus not generating or showing quantum coherence for all times $t$. Based on this, we were able to determine a bound to the maximal rate of chance of the classical system, the classical quantum speed limit.
    When the quantum evolution beats this classical bound, one has a direct certification of nonclassical speedups due to polarization nonclassicality.

    We applied the formalism to the cross-Kerr interaction, a prominent nonlinear dynamics in classical and quantum polarization optics.
    We were able to prove a persistent nonclassical speedup, scaling as $\mathcal{O}(N^{1/2})$ with the photon number $N$.
    Therefore, we provide a quantitative measure for the quantum speedup of this process beyond classical bounds.
    The analysis of specific trajectories further confirms that the unrestricted evolution clearly deviates from the manifold of classical states, isolating polarization nonclassicality as the origin of the observed enhancement.
    The instantaneous speedups accumulate to overall faster quantum processes, saving time to reach a processing goal compared to a purely classical evolution.
    \TA{We further note that the $\mathcal{O}(\sqrt N)$ enhancement is obtained for an ideal, lossless evolution.}
    \LA{To address this ratio in realistic implementations within our framework, one can restrict the open-system dynamics modeled by the Lindblad evolution \cite{pinske2024separabilitylindbladequationdynamical,ares2024restrictedMonteCarlowave-functionmethod}.}

\begin{acknowledgments}
    
    T.A. is grateful to the PhoQS FUTURE graduate program for financial support. 
    T.A. further acknowledges support for the publication cost by the Open Access Publication Fund of Paderborn University.
    J.S. and L.A. acknowledge funding through the QuantERA project QuCABOoSE.
\end{acknowledgments}
\bibliography{src_new}     
\end{document}